\documentclass[aps,prd,twocolumn,superscriptaddress,showpacs,showkeys,floatfix]{revtex4-1}

\usepackage{graphicx}
\usepackage{amssymb}
\usepackage{amsmath}
\usepackage{bm}

\begin{document}


\title{The shadow of dark matter as a shadow of string theory:\\ 
String origin of the dipole term}



\author{Rainer Dick}
\email{rainer.dick@usask.ca}

\affiliation{Department of Physics and Engineering Physics, 
University of Saskatchewan, 116 Science Place, Saskatoon, Canada SK S7N 5E2}



\begin{abstract} 
We point out that the Kalb-Ramond field can couple to the Ramond and Neveu-Schwarz fields
of superstring theory in a way that can generate a coupling of the
Kalb-Ramond field to dark matter dipole moments. 
Electroweak dipole dark matter could then arise from the Ramond sector
of superstrings with low string scale $M_s\lesssim 120$ TeV.
\end{abstract}


\maketitle


\section{Introduction\label{sec:intro}}

After the confirmation of the Higgs particle as the capstone of the Standard Model, solving 
the persistent enigma of dark matter has become a major objective of modern particle physics. 
The completion of the Standard Model (SM) is a great example of successful collaboration between 
theory and experiment, and theoretical models are needed in the ``beyond SM'' era to assist 
with the search for dark matter in particle physics experiments and to provide ideas on what to 
look for. 

It is an intriguing possibility that dark matter might provide a window into low energy 
string phenomenology, e.g.~through the swampland conjecture leading to fading dark 
matter within a tower of light string states \cite{sw1,sw2}, or through models where a low 
string scale \cite{antoniadis1,antoniadis2} implies phenomenologically acceptable dark axions 
or dark gauge bosons \cite{kiritsis,luis,berenstein,honecker}.
It has been pointed out recently \cite{adrd} that electroweak dipole dark matter \cite{kris,PS} 
may open a window into low energy string phenomenology through the Kalb-Ramond field. The
question of the relevant string scale was not addressed in Ref.~\cite{adrd} because no direct
link between the dark matter coupling and mass to the string scale was established. However,
it turns out that the Kalb-Ramond field can couple to the fermionic Ramond fields in such a way 
that a low energy coupling of the Kalb-Ramond field to dark fermions can be traced back to string 
theory. This then allows for an estimate of the maximal string scale for which perturbatively
coupled dipole dark matter could be induced from string theory.

Kalb and Ramond had noticed that gauge interactions between strings can be described in 
analogy to electromagnetic interactions if the basic Nambu-Goto action is amended with a 
coupling term to an anti-symmetric tensor field \cite{KR0}. We follow the conventions 
of Ref.~\cite{adrd} and denote the Kalb-Ramond field with $C_{\mu\nu}=-\,C_{\nu\mu}$ 
to avoid confusion with the electroweak $U_Y(1)$ field strength tensor $B_{\mu\nu}$.
The action for strings coupling to the Kalb-Ramond field is
\begin{eqnarray}\nonumber
S_1&=&-\,T_s\int\! d^2\sigma
\sqrt{(\dot{X}\cdot X')^2-\dot{X}^2 X'^2}
\\ \nonumber
&&+\,\frac{\mu_s}{2}\int\! d^2\sigma\left(\dot{X}^\mu X'^\nu
-\dot{X}^\nu X'^\mu\right)C_{\mu\nu}(X)
\\ \label{eq:action1}
&&+\,g_b\!\int\! d\tau\left[\dot{X}^\mu \mathcal{B}_{\mu}(X)\right]_{\sigma=0}^{\sigma=\ell}
\end{eqnarray}
Here $T_s$ is the string tension and $\mu_s$ is a string charge
with the dimension of mass. The world sheet measure is $d^2\sigma=d\tau d\sigma=d\sigma^+ d\sigma^-/2$,
where $\sigma^\pm=\sigma\pm\tau$.

The string equations of motion which follow from (\ref{eq:action1})
are invariant under the KR gauge symmetry 
\begin{equation}\label{eq:KRg1}
C_{\mu\nu}\to C_{\mu\nu}+\partial_\mu f_\nu-\partial_\nu f_\mu,\quad
\mathcal{B}_\mu\to\mathcal{B}_\mu+(\mu_s/g_b)f_\mu,
\end{equation}
and under the $U(1)$ gauge transformation $\mathcal{B}_\mu\to\mathcal{B}_\mu +\partial_\mu f$,
and these gauge symmetries are preserved through addition of a field theory action for the 
gauge fields,
\begin{eqnarray}\nonumber
S_2&=&\int\!d^4x\left(
-\,\frac{1}{6}C^{\mu\nu\rho}C_{\mu\nu\rho}-\frac{1}{4}\mathcal{B}^{\mu\nu}\mathcal{B}_{\mu\nu}
\right.
\\ \label{eq:kinfields}
&&+\left.\frac{\mu_s}{2g_b}C^{\mu\nu}\mathcal{B}_{\mu\nu}
-\frac{\mu_s^2}{4g_b^2}C^{\mu\nu}C_{\mu\nu}\right)\!,
\end{eqnarray}
where 
\begin{equation}\label{eq:dC}
C_{\mu\nu\rho}=\partial_\mu C_{\nu\rho}+\partial_\nu C_{\rho\mu}+\partial_\rho C_{\mu\nu}
\end{equation}
are the components of the 3-form field strength $C_3=dC$ of the Kalb-Ramond field.
This motivated the proposal in \cite{adrd} to induce a $U_Y(1)$ portal to dark matter
through the Lagrangian
\begin{eqnarray}\nonumber
\mathcal{L}_{BC\chi}&=&\overline{\chi}\left(\mathrm{i}\gamma^\mu\partial_\mu
-m_\chi\right)\chi-\frac{1}{6}C^{\mu\nu\rho}C_{\mu\nu\rho}
\\  \nonumber
&&-\,\frac{1}{2}m^2_C C_{\mu\nu}C^{\mu\nu}
-g_{BC}m_C B^{\mu\nu}C_{\mu\nu}
\\  \label{eq:L2}
&&-\,g_{C\chi}\overline{\chi}S^{\mu\nu}(a_m+\mathrm{i}a_e\gamma_5)\chi C_{\mu\nu}.
\end{eqnarray}
Elimination of $C_{\mu\nu}$ for energies much smaller than $m_C$ generates
electroweak dipole moments for the dark matter field $\chi$,
\begin{eqnarray} \nonumber
\mathcal{L}_{B\chi}&=&\frac{g_{BC}g_{C\chi}}{m_C}B_{\mu\nu}
\overline{\chi}S^{\mu\nu}(a_m+\mathrm{i}a_e\gamma_5)\chi
\\ \nonumber
&=&\frac{g_{BC}g_{C\chi}}{m_C}(F_{\mu\nu}\cos\theta-Z_{\mu\nu}\sin\theta)
\\ \label{eq:ps1}
&&\times\overline{\chi}S^{\mu\nu}(a_m+\mathrm{i}a_e\gamma_5)\chi.
\end{eqnarray}
However, while the bosonic terms in (\ref{eq:L2}) could be motivated 
through the couplings of string world sheets to Kalb-Ramond fields,
the coupling of $C_{\mu\nu}$ to the dark fermions had to be postulated.
The purpose of the present paper is to point out that the
Kalb-Ramond field can couple to the Ramond \cite{ramond}
and Neveu-Schwarz \cite{NS} fields of superstrings
in a way that does induce magnetic dipole couplings at low energies,
thus lending more credibility to the proposal that direct search for 
electroweak dipole dark matter can provide a window into 
low energy string phenomenology.

The possible string origin of the coupling of the Kalb-Ramond field 
to dark dipole moments is introduced in Sec.~\ref{sec:origin} and
conclusions are formulated in Sec.~\ref{sec:conc}. The mapping between
world-sheet spinors and half-differentials is reviewed in Appendix A,
since it helps to understand the proposed coupling of the Kalb-Ramond field
to the half-differentials $\Psi^\mu_{\sqrt{\pm}}$. The Kalb-Ramond field in
Lorentz gauge and in Coulomb gauge and its polarization is reviewed in 
Appendix B.

\section{String origin of magnetic dipole terms}\label{sec:origin}

We wish to provide a string explanation for the coupling of the Kalb-Ramond field 
to the dark fermion dipole moments in Eq.~(\ref{eq:L2}). This will require a coupling
to the Ramond fields $\psi^\mu(\tau,\sigma)$ on the world sheet, which we write as 
components $\Psi^\mu_{\sqrt{\pm}}(\tau,\sigma)$ of half-differentials.
The mapping between 2D spinors and half-differentials is reviewed in Appendix A.

The mapping should respect world-sheet and target space symmetries under coordinate 
and Lorentz transformations. Since 2D spinors are completely equivalent to
half-differentials, and since we can only use $C_{\mu\nu}$ or $C_{\mu\nu\rho}$ for 
the coupling, the unique solution to this problem is
\begin{eqnarray}\nonumber
S_3&=&g_s\int\! d\sigma^+ d\sigma^-\,
C_{\mu\nu\rho}(X)\!\left(\Psi^\mu_{\sqrt{+}}\Psi^\nu_{\sqrt{+}}\partial_- X^\rho\right.
\\ \label{eq:S3}
&&-\left.\Psi^\mu_{\sqrt{-}}\Psi^\nu_{\sqrt{-}}\partial_+ X^\rho\right)\!,
\end{eqnarray}
with a dimensionless coupling constant $g_s$. The minus sign is required by parity invariance
of the world-sheet Lagrangian. 

The coupling term (\ref{eq:S3}) is in world-sheet spinor notation (with metric 
determinant $g\equiv g(\tau,\sigma)$ and zweibein 
components $e^\alpha{}_{a}\equiv e^\alpha{}_a(\tau,\sigma)$) given by
\begin{equation} \label{eq:S3b}
S_3=-\,g_s\int\! d^2\sigma\,\sqrt{-g}\,
C_{\mu\nu\rho}(X) e^\alpha{}_{a}\overline{\psi}^\mu\gamma^a\psi^\nu
\partial_\alpha X^\rho.
\end{equation}

The coupling (\ref{eq:S3}) preserves the KR gauge symmetry $C\to C+df$ (\ref{eq:KRg1}) 
because only the 3-form field strength $dC$ (\ref{eq:dC}) appears. However, for later 
reference it is also useful to explicitly note this on the level of the equations of motion.
The contribution from (\ref{eq:S3}) to the equation of motion of the Kalb-Ramond tensor is
a source current
\begin{eqnarray}\nonumber
J_3^{\mu\nu}(x)&=&-\,\frac{\delta S_3}{\delta C_{\mu\nu}(x)}
=g_s\int\! d\sigma^+ d\sigma^-\,\frac{\partial}{\partial x^\rho}\delta(x-X)
\\ \nonumber
&&\times\!
\left[
\Psi^\rho_{\sqrt{+}}\Psi^\mu_{\sqrt{+}}\partial_- X^\nu 
+\Psi^\nu_{\sqrt{+}}\Psi^\rho_{\sqrt{+}}\partial_- X^\mu 
\right.
\\ \nonumber
&&+\,\Psi^\mu_{\sqrt{+}}\Psi^\nu_{\sqrt{+}}\partial_- X^\rho
-\Psi^\rho_{\sqrt{-}}\Psi^\mu_{\sqrt{-}}\partial_+ X^\nu 
\\ \label{eq:J3}
&&-\left.\Psi^\nu_{\sqrt{-}}\Psi^\rho_{\sqrt{-}}\partial_+ X^\mu
-\Psi^\mu_{\sqrt{-}}\Psi^\nu_{\sqrt{-}}\partial_+ X^\rho
\right]\!.
\end{eqnarray}
This current satisfies 
\begin{equation}\label{eq:dJ3}
\partial_\mu J_3^{\mu\nu}(x)=0
\end{equation}
due to the anti-symmetry of the bracket in (\ref{eq:J3}) under $\mu\leftrightarrow\rho$.
Eq.~(\ref{eq:dJ3}) ensures that the addition $J^{\mu\nu}\to J^{\mu\nu}+J_3^{\mu\nu}$ to the
source terms of the original Kalb-Ramond equations of motion complies with gauge invariance,
see also Appendix B.

The couplings (\ref{eq:action1},\ref{eq:S3}) of the Kalb-Ramond field break the 
world-sheet supersymmetry of the free string theory. However, they do preserve 
the GSO projection \cite{GSO}, and that is what we really need. 
The couplings (\ref{eq:action1},\ref{eq:S3}) correspond to interactions between 
string world sheets which are mediated by a spacetime field.
This leads to a Dyson representation for the scattering matrix,
\begin{equation}\label{eq:Smatrix1}
S_{fi}=\langle f|\mathrm{T}\exp\!\left(-\,\mathrm{i}\int\!d^4x\,\mathcal{H}_I(x)\right)|i\rangle,
\end{equation}
with the Hamiltonian density in the interaction picture.
The contribution to $\mathcal{H}_I$ from Eq.~(\ref{eq:S3}) is in Lorentz 
gauge \cite{note1}
\begin{eqnarray}\nonumber
\mathcal{H}_3(x)&=&-\,g_s\int\! d\sigma^+ d\sigma^-\,
C_{\mu\nu\rho}(x)\!\left(\Psi^\mu_{\sqrt{+}}\Psi^\nu_{\sqrt{+}}\partial_- X^\rho\right.
\\ \label{eq:H3}
&&-\left.\Psi^\mu_{\sqrt{-}}\Psi^\nu_{\sqrt{-}}\partial_+ X^\rho\right)\!
\delta(x-X(\tau,\sigma)).
\end{eqnarray}
The Kalb-Ramond type interactions (\ref{eq:action1},\ref{eq:S3}) extend the
sector of string interactions by introducing a spacetime field which probes 
the bosonic and fermionic string excitations across the world sheet. We could
think of them in terms of vertex operator insertions on the world
sheet, but they do not generate additional constraints from branch cuts.
 The consistent string theories are therefore still built from tensor
products of the Ramond sectors $(\mathrm{R},\pm)$ and the Neveu-Schwarz
sector $(\mathrm{NS},+)$ \cite{polchinski}, where the signs denote the level 
of fermionic world sheet excitations modulo 2. The interactions 
(\ref{eq:action1},\ref{eq:S3}) can change the level 
of fermionic world sheet excitations only in even steps, and they cannot
change periodicity conditions on closed strings or boundary counditions
on open strings. Therefore they map states in a consistent string theory
into states of that theory. As a consequence, and in keeping with the proposal 
of Kalb and Ramond of fundamental 1-branes interacting through exchange 
of fundamental 0-branes, the states of string theory with Kalb-Ramond type 
interactions are tensor products of string states in the Ramond or 
Neveu-Schwarz sectors of the theory with Kalb-Ramond 
states $|\bm{k}_1,\ldots\bm{k}_M;\bm{p}_1,\ldots\bm{p}_N,\alpha_1,\ldots\alpha_N\rangle$, 
where $\bm{k}_I$ and $\bm{p}_I$ are momenta of Kalb-Ramond tensor fields $C_{\mu\nu}$
and vector fields $\mathcal{B}_\mu$, respectively, and the $\alpha_I$
denote the polarizations of the vector fields \cite{note2}.
There are no polarization labels for the tensor field, because anti-symmetric 
tensor fields $C_{\mu\nu}$ with gauge symmetry and 
3-momentum $\bm{k}$ have only one possible 
physical polarization state in four dimensions, \textit{viz.}
 $\epsilon^{(1)}_\mu(\bm{k})\epsilon^{(2)}_\nu(\bm{k})
-\bm{\epsilon}^{(2)}_\mu(\bm{k})\epsilon^{(1)}_\nu(\bm{k})$, 
where $k\cdot\epsilon^{(\alpha)}(\bm{k})=\bm{k}\cdot\bm{\epsilon}^{(\alpha)}(\bm{k})=0$,
i.e.~the basis $\epsilon^{(\alpha)}(\bm{k})$ is like the polarization basis of 
physical states of a massless vector field with momentum $\bm{k}$. However, 
in Lorentz gauge there are two additional unphysical polarizations of the form 
 $\epsilon^{(\alpha)}_\mu(\bm{k})\epsilon^{(3)}_\nu(\bm{k})
-\epsilon^{(3)}_\mu(\bm{k})\epsilon^{(\alpha)}_\nu(\bm{k})$, 
$k\cdot\epsilon^{(3)}(\bm{k})=0\neq\bm{k}\cdot\bm{\epsilon}^{(3)}(\bm{k})$,
see also Appendix B.

If we evaluate the expression (\ref{eq:S3}) on the low-energy states
\begin{equation}\label{eq:LowER1}
|\bm{k},s\rangle=|0\rangle u(\bm{k},s),\quad
u^+(\bm{k},s)\cdot u(\bm{k},s')=\delta_{s,s'},
\end{equation}
\begin{equation}\label{eq:LowER2}
\partial_\pm X^\mu\Rightarrow\pm k^\mu/2k^0,
\quad \Psi^\mu_{\sqrt{\pm}}\Rightarrow\mathrm{i}\gamma^\mu/\sqrt{8},
\end{equation} 
of the Ramond sector, we find in temporal gauge $\tau=t$,
$\int d\sigma=\ell=k^0/T_s$,
\begin{eqnarray}\nonumber
S_3&=&\frac{g_s}{8}\int\! dt\,
C_{\mu\nu\rho}\!\left(x+\frac{kt}{k^0}\right)\frac{k^\rho}{T_s}
\\ \label{eq:S3c}
&&\times \overline{u}(\bm{k},s)\cdot[\gamma^\mu,\gamma^\nu]\cdot u(\bm{k},s).
\end{eqnarray}
This describes the interaction of the Kalb-Ramond field with a classical fermion. 
The corresponding quantum mechanical action is
\begin{equation} \label{eq:S3d}
S_3=g_s\frac{\pi^3}{\mathrm{i}T_s}\int\! d^4x\,C^{\mu\nu\rho}(x)
\overline{\chi}(x)\cdot[\gamma_\mu,\gamma_\nu]\cdot\partial_\rho\chi(x).
\end{equation}
The non-relativistic limit of Eq.~(\ref{eq:S3d}) corresponds to a coupling
\begin{equation} \label{eq:S3e}
S_3=-\,4g_s\frac{\pi^3}{T_s}m_\chi m_C\int\! d^4x\,C^{ij}(x)
\overline{\chi}(x) S_{ij}\chi(x),
\end{equation}
where we invoked the standard assumption that the low energy string states in 
the \textit{a priori} massless sector acquire masses through symmetry breaking.

The coupling of the Kalb-Ramond field to fermions on the world-sheet (\ref{eq:S3})
and in the resulting field theory coupling (\ref {eq:S3d}) are explicitly invariant
under the KR gauge symmetry because the fermions couple to the Kalb-Ramond field
strength. The resulting coupling (\ref{eq:S3e}) in the non-relativistic limit
is effectively invariant within that limit due to $m_C\gg|\bm{k}|$, where $\bm{k}$
is the 3-momentum of the Kalb-Ramond field. We 
have $m_C |C_{ij}|\simeq|\partial_0 C_{ij}|\gg|\partial_i C_{0j}|$, 
and therefore $m_C C_{ij}\simeq\mathrm{i}C_{0ij}$.

Elimination of $C^{ij}(x)$ from the action with the terms (\ref{eq:L2}) 
yields a magnetic dipole coupling with coupling constant
\begin{equation}\label{eq:Ts1}
\frac{a_m}{M_d}=g_{BC}g_{C\chi}\frac{a_m}{m_C}=4g_sg_{BC}\frac{\pi^3}{T_s}m_\chi.
\end{equation}
However, the coupling scale $M=M_d/a_m$ can be determined as a function 
of dark matter mass $m_\chi$ through the requirement of thermal dark matter
creation in the early universe. This led to $M\simeq 23$ PeV for $m_\chi=1$ MeV 
and $M\simeq 3.7$ TeV for $m_\chi=10$ GeV, or 
roughly \cite{note3} $m_\chi M\simeq 3\times 10^4\,\mathrm{GeV}^2$,
see Fig.~1 in Ref.~\cite{adrd}. However, Eq.~(\ref{eq:Ts1}) relates this
to the string scale,
\begin{equation}\label{eq:Ts2}
T_s=4g_sg_{BC}\pi^3 m_\chi M.
\end{equation}
String-induced electroweak dipole dark matter with perturbative couplings 
would therefore require a low string scale $\sqrt{8\pi T_s}\lesssim 120$ TeV.

\section{Conclusion}\label{sec:conc}

Inclusion of the missing piece (\ref{eq:S3}) in the string derivation of
electroweak dipole dark matter relates the string tension to dark matter
parameters. Assuming perturbative couplings and that dark matter is
created from thermal freeze-out then leads to the 
requirement $M_s\lesssim 120$ TeV. Since string theory appears to be unique
in providing a mechanism to induce electroweak dipole moments in dark matter,
discovery of electroweak dipole dark matter with a mass and recoil cross section 
which agrees with the predictions from thermal dark matter creation would be a 
strong hint for low-scale string theory.

It is intriguing that the geometry of string ground states allows for the
construction of phenomenologically viable models with fundamental string 
scales all the way down to the TeV 
scale \cite{antoniadis1,antoniadis2,kiritsis,luis,berenstein,honecker}.
 Furthermore, it is also conceivable that there may exist different
incarnations of fundamental strings with different tensions $T_s$. In that 
case the Kalb-Ramond model (\ref{eq:action1}) and its extension (\ref{eq:S3}) 
could find a natural ``pure string'' explanation within a hierarchy of string 
scales $T_s\ll T_P=M^2_{\rm Planck}/8\pi$. The Kalb-Ramond field $C_{\mu\nu}(x)$ would 
then correspond to the low energy description of the anti-symmetric tensor fields of 
Planck scale superstrings with tension $T_P$, and the couplings (\ref{eq:action1},\ref{eq:S3}) 
would arise from world-sheet couplings between strings with different tensions. If the low 
scale string sector contains only Ramond sector states, such a scenario would not need to 
invoke any further assumption about the nature of the string ground state to explain 
why $M_{\rm Planck}\gg M_s$. Either way, the possibility to induce electroweak dipole dark 
matter from string theory adds to the anticipation that low scale string theory should become 
a leading paradigm for particle physics searches beyond the Standard Model, both in direct 
dark matter search experiments and at future colliders.

\section*{Appendix A: Half-differentials and fermions on the world sheet}

Here we review the mapping between spinors and half-differentials on
world sheets in the conformal gauge
\begin{equation}
g_{\tau\tau}+g_{\sigma\sigma}=\dot{X}^2+X'^2=0,\quad
g_{\tau\sigma}=\dot{X}\cdot X'=0.
\end{equation}
We denote the remaining degree of freedom in the metric by
$\phi(\tau,\sigma)$,
\begin{equation}
g_{\sigma\sigma}=-\,g_{\tau\tau}=\exp(2\phi).
\end{equation}
The metric in the corresponding two-dimensional light-cone coordinates
$\sigma^\pm=\sigma\pm\tau$,
\begin{equation}
g_{++}=g_{--}=0,\quad g_{+-}=\frac{1}{2}\exp(2\phi)
=\frac{1}{2}\sqrt{-g},
\end{equation}
corresponds to zweibein components $e_\alpha{}^a$ (note
$\eta_{+-}=1/2$),
\begin{equation}
e_+{}^+=e_-{}^-=\exp(\phi),\quad e_+{}^-=e_-{}^+=0,
\end{equation}
\begin{equation}
e_{-+}=e_{+-}=\frac{1}{2}\exp(\phi),\quad e_{++}=e_{--}=0.
\end{equation}
The set of coordinate transformations is restricted 
by the conformal gauge \cite{note4} to
\begin{equation}\label{eq:2Ddiffs}
\sigma^+\to\sigma'^+(\sigma^+),\quad\sigma^-\to\sigma'^-(\sigma^-).
\end{equation}
On the other hand, the symmetry group $SO(1,1)$ of tangent plane
Lorentz boosts also factorizes in the light-cone basis of tangent
vectors. A boost with parameter
\begin{equation}
u=\mathrm{artanh}(\beta)
=\frac{1}{2}\ln\!\left(\frac{1+\beta}{1-\beta}\right)
\end{equation}
in the tangent plane yields in the light-cone basis $v^\pm=v^1\pm v^0$ 
the transformation law
\begin{equation}\label{eq:lambda2}
\left(\begin{array}{c}
v'^+\\
v'^-\\
\end{array}
\right)
=\left(
\begin{array}{cc}
\Lambda^+{}_+ & 0\\
0 & \Lambda^-{}_-
\end{array}
\right)\left(\begin{array}{c}
v^+\\
v^-\\
\end{array}
\right).
\end{equation}
with components 
\begin{equation}
\Lambda^+{}_+=\left(\frac{1-\beta}{1+\beta}\right)^{1/2},\quad
\Lambda^-{}_-=\left(\frac{1+\beta}{1-\beta}\right)^{1/2}.
\end{equation}
This is the exact square of the transformation of the spinor components in 
Weyl representation. Every Weyl representation of 2D $\gamma$ matrices yields 
either $\mathrm{i}S_{10}$ or $-\,\mathrm{i}S_{10}$ as the spinor representation 
of the boost generator, where
\begin{equation}\label{eq:S10}
\mathrm{i}S_{10}=\frac{1}{2}\gamma_0\cdot\gamma_1
=\frac{\mathrm{1}}{2}\left(
\begin{array}{cc}
 -1 & \,\,\,\,0 \\
 \,\,\,\,0 & \,\,\,\,1 \\
\end{array}\right)\!.
\end{equation}
The corresponding Lorentz boost on the spinors is
\begin{eqnarray}\nonumber
\left(\begin{array}{c}
\psi'^{\sqrt{+}}\\
\psi'^{\sqrt{-}}\\
\end{array}
\right)
&=&
\left(
\begin{array}{cc}
\exp(-\,u/2) & 0\\
0 & \exp(u/2)\\
\end{array}
\right)\!\left(\begin{array}{c}
\psi^{\sqrt{+}}\\
\psi^{\sqrt{-}}\\
\end{array}
\right)
\\ \label{eq:spintrans}
&=&
\left(
\begin{array}{cc}
\left(\frac{1-\beta}{1+\beta}\right)^{1/4} & 0\\
0 & \left(\frac{1+\beta}{1-\beta}\right)^{1/4} 
\end{array}
\right)\!\left(\begin{array}{c}
\psi^{\sqrt{+}}\\
\psi^{\sqrt{-}}\\
\end{array}
\right)\!,
\end{eqnarray}
i.e.~the matrix elements of the spinor representation relate to the
matrix elements of the vector representation through
\begin{equation}
\Lambda^{\sqrt{\pm}}{}_{\sqrt{\pm}}=\sqrt{\Lambda^{\pm}{}_{\pm}}.
\end{equation}
This motivates the assignments of indices to the components of the 2D Dirac 
spinor in the Weyl basis: $(\psi^{\sqrt{+}})^2$ and 
$(\psi^{\sqrt{-}})^2$ transform like the components of a tangent vector 
in the light-cone basis. 

To be specific we use in the following calculations the real Weyl 
representation of 2D $\gamma$ matrices
\begin{equation}\label{eq:2dgamma}
\gamma^0=\left(
\begin{array}{cc}
 \,\,\,\,\,0 & -1\\
 -1 & \,\,\,\,\,0\\
\end{array}
\right),\quad
\gamma^1=\left(
\begin{array}{cc}
 \,\,\,\,\,0 & \,\,\,\,1\\
 -1 & \,\,\,\,0\\
\end{array}
\right)\!.
\end{equation}
However, the mapping between world-sheet spinors and half-differentials, 
and the fermion action in terms of half-differentials, follow in 
the same form for every Weyl representation.

The tangent plane gamma matrices in the light
cone basis follow as
\begin{equation}\label{eq:gamma+-}
\gamma^+=-\,2\left(
\begin{array}{cc}
0 & \,\,\,\,0\\
1 & \,\,\,\,0\\
\end{array}
\right),
\quad
\gamma^-=2\left(
\begin{array}{cc}
0 & \,\,\,\,1\\
0 & \,\,\,\,0\\
\end{array}
\right),
\end{equation}
and the transformation of the integration measure to the
light-cone coordinates on the world sheet is
\begin{equation}
d\tau d\sigma\sqrt{-g}=d\sigma^+ d\sigma^- g_{+-}.
\end{equation}
The curved space generalization
of the kinetic terms in the Dirac action is
\begin{align}\nonumber
&\frac{\mathrm{i}}{2}
e^\alpha{}_a\!\left[(\overline{\psi}\Omega_\alpha
-\partial_\alpha\overline{\psi})\gamma^a\psi
+\overline{\psi}\gamma^a
(\partial_\alpha\psi+\Omega_\alpha\cdot\psi)\right]
\\
&=\frac{\mathrm{i}}{2}e^\alpha{}_a\!\left[\overline{\psi}\gamma^a\partial_\alpha\psi
-\partial_\alpha\overline{\psi}\cdot\gamma^a\psi
+\overline{\psi}\{\Omega_\alpha,\gamma^a\}\psi\right]\!.
\end{align}
However, the spin-connection in two dimensions,
\begin{equation}
\Omega_\alpha
=-\,\mathrm{i}\Gamma^0{}_{1\alpha}S^1{}_0
=\frac{1}{2}\Gamma^0{}_{1\alpha}\gamma^1\gamma_0,
\end{equation}
anti-commutes both with $\gamma^0$ and $\gamma^1$, $\{\Omega_\alpha,\gamma^a\}=0$.
Therefore, splitting the derivative symmetrically between
$\overline{\psi}$ and $\psi$ cancels the spin connection
from the Dirac action in two dimensions.  
The resulting action combining the Fubini-Veneziano
fields and Dirac spinors on the world sheet is
\begin{eqnarray} \nonumber
S&=&-\int\!d\tau\int_0^{\ell}\! d\sigma\,\sqrt{-g}\left[\frac{T_s}{2}
g^{\alpha\beta}\partial_\alpha X\cdot\partial_\beta X\right.
\\ \nonumber
&&+\left.\frac{\mathrm{i}}{2}\sqrt{T_s}e^\alpha{}_a\left(
\partial_\alpha\overline{\psi}\cdot\gamma^a\cdot\psi
-\overline{\psi}\cdot\gamma^a\cdot\partial_\alpha\psi\right)
\right]
\\ \nonumber
&=&-\int\!d\sigma^+ d\sigma^- g_{+-}\Big[T_s
g^{+-}\partial_+ X\cdot\partial_- X
\\ \nonumber
&&+\,\frac{\mathrm{i}}{2}\sqrt{T_s}e^+{}_+\left(
\partial_+\overline{\psi}\cdot\gamma^+\cdot\psi
-\overline{\psi}\cdot\gamma^+\cdot\partial_+\psi\right)
\\ \label{gsaction1}
&&
+\,\frac{\mathrm{i}}{2}\sqrt{T_s}e^-{}_-\left(
\partial_-\overline{\psi}\cdot\gamma^-\cdot\psi
-\overline{\psi}\cdot\gamma^-\cdot\partial_-\psi\right)\Big].
\end{eqnarray}
Substituting the spinor components $\psi=(\psi^{\sqrt{+}},\psi^{\sqrt{-}})$,
$\overline{\psi}=(-\,\psi^{\sqrt{-},*},-\,\psi^{\sqrt{+},*})$,
and the $\gamma$-matrices (\ref{eq:gamma+-}) yields
\begin{eqnarray}\nonumber
S&=&-\int\!d\sigma^+ d\sigma^-\Big[
T_s\partial_+ X\cdot\partial_- X
\\ \nonumber
&&+\,\mathrm{i}\sqrt{T_s}e_{-+}\left(
\partial_+\psi^{\sqrt{+},*}\cdot\psi^{\sqrt{+}}
-\psi^{\sqrt{+},*}\cdot\partial_+\psi^{\sqrt{+}}\right)
\\ \nonumber
&&
+\,\mathrm{i}\sqrt{T_s}e_{+-}\left(
\psi^{\sqrt{-},*}\cdot\partial_-\psi^{\sqrt{-}}-
\partial_-\psi^{\sqrt{-},*}\cdot\psi^{\sqrt{-}}\right)\Big]
\\ \nonumber
&=&-\!\int\!d\sigma^+ d\sigma^-\Big[
T_s\partial_+ X\cdot\partial_- X
\\ \nonumber
&&+\,\mathrm{i}\sqrt{T_s}
\partial_+\left(\sqrt{e_{-+}}\psi^{\sqrt{+},*}\right)\cdot\left(\sqrt{e_{-+}}\psi^{\sqrt{+}}\right)
\\ \nonumber
&&-\,\mathrm{i}\sqrt{T_s}\left(\sqrt{e_{-+}}\psi^{\sqrt{+},*}\right)
\cdot\partial_+\left(\sqrt{e_{-+}}\psi^{\sqrt{+}}\right)
\\ \nonumber
&&+\,\mathrm{i}\sqrt{T_s}\left(
\sqrt{e_{+-}}\psi^{\sqrt{-},*}\right)\cdot\partial_-\left(\sqrt{e_{+-}}\psi^{\sqrt{-}}\right)
\\ \nonumber
&&-\,\mathrm{i}\sqrt{T_s}\partial_-\left(\sqrt{e_{+-}}\psi^{\sqrt{-},*}\right)
\cdot\left(\sqrt{e_{+-}}\psi^{\sqrt{-}}\right)
\Big].
\end{eqnarray}
Combining the spinor and zweibein components yields the half-differentials
\begin{eqnarray}\label{eq:Psidef1}
\Psi_{\sqrt{-}}(\xi)&=&\sqrt{e_{-+}}(\xi)\psi^{\sqrt{+}}(\xi),
\\ \label{eq:Psidef2}
\Psi_{\sqrt{+}}(\xi)&=&\sqrt{e_{+-}}(\xi)\psi^{\sqrt{-}}(\xi),
\end{eqnarray}
and the action
\begin{eqnarray}\nonumber
S&=&-\!\int\!d\sigma^+ d\sigma^-\Big[T_s\partial_+ X\cdot\partial_- X
\\ \nonumber
&&
+\,\mathrm{i}\sqrt{T_s}
\Big(\Psi_{\sqrt{+}}^*\cdot\partial_-\Psi_{\sqrt{+}}
-\partial_-\Psi_{\sqrt{+}}^*\cdot\Psi_{\sqrt{+}}\Big)
\\ \label{eq:gsaction2}
&&
+\,\mathrm{i}\sqrt{T_s}
\Big(\partial_+\Psi_{\sqrt{-}}^*\cdot\Psi_{\sqrt{-}}
-\Psi_{\sqrt{-}}^*\cdot\partial_+\Psi_{\sqrt{-}}\Big)\Big].
\end{eqnarray}
The 2D metric has completely disappeared, in the scalar sector due to the 
conformal gauge and in the spinor sector due to absorption into the 
half-differentials. 

The spinors are invariant under coordinate transformations on the world 
sheet and transform under Lorentz transformations in the tangent plane
according to (\ref{eq:spintrans}). The 
half-differentials (\ref{eq:Psidef1},\ref{eq:Psidef2}) are invariant 
under Lorentz transformations, but transform under the coordinate 
transformations (\ref{eq:2Ddiffs}) according to
\begin{eqnarray}\label{eq:Psitrans1}
\Psi'_{\sqrt{-}}(\xi')&=&\Psi_{\sqrt{-}}(\xi)
\sqrt{d\sigma^-/d\sigma'^-},
\\ \label{eq:Psitrans2}
\Psi'_{\sqrt{+}}(\xi')&=&\Psi_{\sqrt{+}}(\xi)
\sqrt{d\sigma^+/d\sigma'^+}.
\end{eqnarray}

The phases of the anti-commuting half-differentials decouple in the 
action (\ref{eq:gsaction2}), in agreement with the real transformation 
behavior (\ref{eq:spintrans}) of the spinors in Weyl representation. 
The standard phase convention is
\begin{equation}\label{eq:realpsi}
\Psi_{\sqrt{\pm}}^{\mu *}=\Psi_{\sqrt{\pm}}^\mu.
\end{equation}
The superconformal transformation parameters $\epsilon^{\sqrt{\pm}}(\sigma^\pm)$
are then real anti-commuting $(-1/2)$-differentials \cite{note5} 
which yield the transformations
\begin{equation}\label{eq:susytrans}
\delta X^\mu=\frac{\mathrm{i}}{\sqrt{T_s}}\left(
\epsilon^{\sqrt{+}}\Psi^\mu_{\sqrt{+}}
+\epsilon^{\sqrt{-}}\Psi^\mu_{\sqrt{-}}\right),
\end{equation}
\begin{equation}\label{eq:susytrans2}
\delta\Psi^\mu_{\sqrt{+}}=-\,\frac{1}{2}
\epsilon^{\sqrt{+}}\partial_+ X^\mu,\quad
\delta\Psi^\mu_{\sqrt{-}}=\frac{1}{2}
\epsilon^{\sqrt{-}}\partial_- X^\mu.
\end{equation}

\section*{Appendix B: The Kalb-Ramond field in Lorentz gauge and in Coulomb gauge,
and its polarizations}

 We used Lorentz gauge for the discussion of the preservation of GSO projection sectors
in Sec.~\ref{sec:origin}. Lorentz gauge is also preferred for covariant perturbation theory
and therefore we also take this opportunity to point out that it can be imposed
simultaneously on both $C_{\mu\nu}$ and $\mathcal{B}_\mu$. After 
imposing $\partial^\mu\mathcal{B}_\mu=0$ in the standard way using the $U(1)$ gauge symmetry
of the vector field, we can use a KR gauge transformation $C^{(0)}_{\mu\nu}\to C_{\mu\nu}$
with the KR gauge function 
\begin{equation}
f_\nu(x)=\int d^4x' G(x-x')\partial'^\mu C^{(0)}_{\mu\nu}(x')
\end{equation}
to impose $\partial^\mu C_{\mu\nu}(x)=0$ while maintaining $\partial^\mu\mathcal{B}_\mu=0$.
The Green function is the retarded function
\begin{equation}
G(x)=\frac{1}{4\pi|\bm{x}|}\delta(t-|\bm{x}|),\quad
\partial^2G(x)=-\,\delta(x).
\end{equation}
This preserves $\partial^\mu\mathcal{B}_\mu=0$ due to
\begin{eqnarray}\nonumber
\partial^\nu f_\nu(x)&=&\int\!d^3\bm{x}'\left.\frac{1}{4\pi|\bm{x}-\bm{x}'|}
\partial'^\mu\partial'^\nu C^{(0)}_{\mu\nu}(x')\right|_{t'=t-|\bm{x}-\bm{x}'|}
\\
&=&0,
\end{eqnarray}
and ensures 
that $C_{\mu\nu}(x)=C^{(0)}_{\mu\nu}(x)+\partial_\mu f_\nu(x)-\partial_\nu f_\mu(x)$
satisfies $\partial^\mu C_{\mu\nu}(x)=0$.

On the other hand, the dominant contribution to electromagnetic interactions
between low-velocity charged particles can be expressed in Coulomb gauge through
Coulomb potentials. In string theory with Kalb-Ramond interactions, Coulomb
gauge is useful to describe the impact of Kalb-Ramond exchange on string-string
interaction forces. Therefore it is also useful to know that KR gauge symmetry
can be used to impose a Coulomb gauge condition $\partial^iC_{i\mu}=0$ for the tensor field
while preserving Coulomb gauge for the vector field. The KR gauge function
\begin{equation}
f_\nu(\bm{x},t)=\int\!d^3\bm{x}'\,\frac{1}{4\pi|\bm{x}-\bm{x}'|}\partial'^i C^{(0)}_{i\nu}(\bm{x}',t)
\end{equation}
accomplishes this due to $\partial^i f_i(x)=0$.

To determine the polarizations of the Kalb-Ramond tensor in the interaction picture states,
we note that
\begin{equation}
\partial_\mu C^{\mu\nu\rho}-m_C^2 C^{\nu\rho}=-\,J^{\nu\rho},
\end{equation}
where the currents $J^{\nu\rho}$ satisfy the consistency condition
\begin{equation}\label{eq:consistency}
\partial_\nu J^{\nu\rho}=m_C^2\partial_\nu  C^{\nu\rho}
\end{equation}
as a consequence
of the contributions of the vector field $\mathcal{B}_\mu$
which maintains the KR gauge symmetry in the presence of the mass 
term for $C_{\mu\nu}$, see Eqs.~(10-13) in Ref.~\cite{adrd}
and Eqs.~(\ref{eq:J3},\ref{eq:dJ3}) in Sec.~\ref{sec:origin}.
The relevant mode expansions of the interaction picture field operators 
therefore still contain only one physical polarization state \cite{OP}.
Coulomb's law and the mode expansion for the remaining degrees of freedom
in Coulomb gauge are
\begin{equation}
(\Delta-m_C^2)C^{0k}=-\,J^{0k}
\end{equation}
and
\begin{eqnarray} \nonumber
C_{ij}(x)&=&\int\!\frac{d^3\bm{k}}{4\sqrt{\pi^3E(\bm{k})}}\,\hat{k}^h\epsilon_{hij}
[a(\bm{k})\exp(\mathrm{i}k\cdot x)
\\ \label{eq:CGmodes}
&&+\,a^+(\bm{k})\exp(-\,\mathrm{i}k\cdot x)],
\end{eqnarray}
respectively, whereas the mode expansion in Lorentz gauge is
\begin{eqnarray} \nonumber
C_{\mu\nu}(x)&=&\int\!\frac{d^3\bm{k}}{4\sqrt{\pi^3E(\bm{k})}}\,\epsilon_{\alpha\beta\gamma}
\epsilon^{(\beta)}_\mu(\bm{k})\epsilon^{(\gamma)}_\nu(\bm{k})
\\ \nonumber
&&\times [a^{(\alpha)}(\bm{k})\exp(\mathrm{i}k\cdot x)
\\ \label{eq:LGmodes}
&&+\,a^{(\alpha)+}(\bm{k})\exp(-\,\mathrm{i}k\cdot x)].
\end{eqnarray}
We choose polarization vectors $\epsilon^{(\alpha)}_\mu(\bm{k})$
such that for $\alpha\in\{1,2\}$
\begin{equation}
k\cdot\epsilon^{(\alpha)}(\bm{k})=\bm{k}\cdot\bm{\epsilon}^{(\alpha)}(\bm{k})=0,
\end{equation}
whereas $\epsilon_0^{(3)}(\bm{k})\neq 0$,
\begin{equation}
k\cdot\epsilon^{(3)}(\bm{k})=0\neq\bm{k}\cdot\bm{\epsilon}^{(3)}(\bm{k}).
\end{equation}
Comparison with the operator (\ref{eq:CGmodes}) in Coulomb gauge shows 
that the single transverse physical polarization in Lorentz gauge is
 $\epsilon^{(1)}_\mu(\bm{k})\epsilon^{(2)}_\nu(\bm{k})
-\epsilon^{(2)}_\mu(\bm{k})\epsilon^{(1)}_\nu(\bm{k})$,
\begin{equation}
[\bm{\epsilon}^{(1)}(\bm{k})\otimes\bm{\epsilon}^{(2)}(\bm{k})
-\bm{\epsilon}^{(2)}(\bm{k})\otimes\bm{\epsilon}^{(1)}(\bm{k})]\cdot\bm{k}=0,
\end{equation}
whereas the two spatially longitudinal polarizations,
 \begin{equation}
[\bm{\epsilon}^{(\alpha)}(\bm{k})\otimes\bm{\epsilon}^{(3)}(\bm{k})
-\bm{\epsilon}^{(3)}(\bm{k})\otimes\bm{\epsilon}^{(\alpha)}(\bm{k})]_{\alpha\in\{1,2\}}
\cdot\bm{k}\neq 0,
\end{equation}
are unphysical.
Therefore the physical states of the Kalb-Ramond tensor,
\begin{equation}
|\bm{k}_1,\ldots\bm{k}_N\rangle=a^{(3)+}(\bm{k}_1)\ldots a^{(3)+}(\bm{k}_N)|0\rangle,
\end{equation}
look like spin-0 states. However, it is not possible to consistently replace
$C_{\mu\nu}$ in interaction terms through a duality transformation into an axion.
At the classical level, a relation of 
the form $C_{\mu\nu\rho}=\epsilon_{\mu\nu\rho\sigma}\partial^\sigma a$ would require 
both fields to be massless free fields, and even if the equation would
be consistent at the classical level, it would not constitute a canonical
transformation of quantum operators.

\begin{acknowledgments} 
This research was supported at the University of Saskatchewan
through a President's NSERC grant.
\end{acknowledgments}


\begin{thebibliography}{88}
\bibitem{note1}
Lorentz gauge is not essential, of course, but convenient. 
Just like in QED, it yields the Lagrangian interaction terms as covariant vertices. 
It can be imposed on $C^{\mu\nu}$ using the KR gauge symmetry (\ref{eq:KRg1})
while preserving Lorentz gauge for $\mathcal{B}_\mu$, see Appendix B.
\bibitem{note2}
There is also another possibility. In multi-scale string theories with different
kinds of strings with different tensions $T_s$, $T_{s'}$, the Kalb-Ramond picture
might correspond to an effective intermediate energy theory for energies
$T_s\lesssim E\ll T_{s'}$, such that for the high scale string modes  an effective field 
theory description applies, whereas for the low-scale strings the world sheet
description is still needed. In that case, the Kalb-Ramond field could arise from
the field theory description of the anti-symmetric tensor modes of the high tension 
string theory. The states in the intermediate energy effective theory would still
have the form described in the text, but the Kalb-Ramond factors would arise from
underlying string states in the high-tension string theory.
\bibitem{note3}
This is a generic feature of thermal dark matter creation.
Larger dark matter mass requires stronger coupling for later freeze-out, since
larger $m_\chi$ requires fewer dark matter particles to explain the observed
dark matter abundance.
\bibitem{note5}
The corresponding Majorana spinor components are 
$\tilde{\epsilon}^{\sqrt{\pm}}=\sqrt{e^{+-}g_{+-}}\epsilon^{\sqrt{\pm}}=\exp(\phi/2)\epsilon^{\sqrt{\pm}}$.
\end{thebibliography}
\end{document}